\documentclass[showpacs]{revtex4}
\usepackage{epsfig}
\def\beq{\begin{equation}}
\def\enq{\end{equation}}
\def\beqa{\begin{eqnarray}}
\def\enqa{\end{eqnarray}}

\def\MeV{\nobreak\,\mbox{MeV}}
\def\GeV{\nobreak\,\mbox{GeV}}

\def\qq{\lag\bar{q}q\rag}

\def\Gd{\lag g^2G^2\rag}
\def\G3{\lag g^3G^3\rag}

\def\rh{\rho}
\def\si{\sigma}

\def\al{\alpha}
\def\be{\beta}
\def\alma{\alpha_{max}}
\def\almi{\alpha_{min}}
\def\bemi{\beta_{min}}
\def\mme{m_{D^*D^*}}
\def\mmo{m_{D_1D}}
\def\lb{\label}
\def\nn{\nonumber}

\newcommand{\rag}{\rangle}
\newcommand{\lag}{\langle}

\begin{document}

\title{\sc Can the $\pi^+\chi_{c1}$ resonance structures be $D^*\bar{D}^*$
and $D_1\bar{D}$ molecules?}
\author{Su Houng Lee}
\email{suhoung@phya.yonsei.ac.kr}
\affiliation{Institute of Physics and Applied Physics, Yonsei University,
Seoul 120-749, Korea}
\author{Kenji Morita}
\email{morita@phya.yonsei.ac.kr}
\affiliation{Institute of Physics and Applied Physics, Yonsei University,
Seoul 120-749, Korea}
\author{Marina Nielsen}
\email{mnielsen@if.usp.br}
\affiliation{Instituto de F\'{\i}sica, Universidade de S\~{a}o Paulo,
C.P. 66318, 05389-970 S\~{a}o Paulo, SP, Brazil}

\begin{abstract}
We use QCD  sum rules to study the recently observed resonance-like
structures in the $\pi^+\chi_{c1}$ mass distribution, $Z_1^+(4050)$ and
$Z_2^+(4250)$, considered as $D^{*+}\bar{D}^{*0}$ and $D_1^+\bar{D}^0+
D^+\bar{D}_1^0$ molecules with the quantum number $J^P=0^+$ and $J^P=1^-$
respectively.
We consider the contributions of condensates up to dimension eight and
work at leading order in $\alpha_s$. We obtain $m_{D^*D^*}=(4.15\pm0.12)~
\GeV$, around 100 MeV above the $D^*D^*$ threshold, and $m_{D_1D}=(4.19\pm
0.22)~\GeV$, around 100 MeV below the $D_1D$ threshold.  We conclude that
the $D^{*+}\bar{D}^{*0}$ state is probably a virtual state that is not
related with the $Z_1^+(4050)$ resonance-like structure. In the case of the
$D_1D$ molecular state, considering the errors, its mass is consistent
with both $Z_1^+(4050)$ and $Z_2^+(4250)$ resonance-like structures. 
Therefore, we conclude that no definite conclusion can be drawn for this 
state from the present analysis.
\end{abstract}

\maketitle


The recent discovery of several missing states and a number of unexpected
charmonium like resonances in B-factories has revitalized the interest in
the espectroscopy of the charmonium states. There is growing evidence that
at least some of these new states are non conventional $c\bar{c}$ states,
such as mesonic molecules, tetraquarks, and/or hybrid mesons.
Among these new mesons, some have their masses
very close to the meson-meson threshold like the $X(3872)$ \cite{belle1}
and the $Z^+(4430)$ \cite{belle2}. Of special importance is the
appearance of the $Z^+(4430)$, observed in the $\pi^+\psi^\prime$ mass
spectrum produced in the $\bar{B}^0\to K^-\pi^+\psi^\prime$ decays.
Being a charged state it can not be described as ordinary $c\bar{c}$
meson. Its nature is completely open, but an intriguing possibility is the
interpretation as tetraquark state or molecular state \cite{swan,meng,lee}.

The $Z^+(4430)$ observation motivated studies of other $\bar{B}^0\to K^-\pi^+
(c\bar{c})$ decays. In particular, the Belle Collaboration has recently
reported the observation of two resonance-like structures in the $\pi^+
\chi_{c1}$ mass distribution \cite{belle3}. The significance of each of the
$\pi^+\chi_{c1}$ structures exceeds 5$\sigma$ and, if they are interpreted
as meson states, their minimal quark content must be $c\bar{c}u\bar{d}$.
They were called $Z_1^+(4050)$ and $Z_2^+(4250)$, and their masses and widths
are $M_1=(4051\pm14^{+20}_{-41})~\MeV$, $\Gamma_1=82^{+21+47}_{-17-22}~\MeV$,
$M_2=(4248^{+44+180}_{-29-~35})~\MeV$, $\Gamma_2=177^{+54+316}_{-39-~61}~
\MeV$.

There are already theoretical interpretations for these structures
as tetraquark states with $J^P=1^-$ \cite{wang} and as molecular
$D^*\bar{D}^*$ state with $J^P=0^+$ \cite{llz}. In this work, due to the
closeness of the $Z_1^+(4050)$ and $Z_2^+(4250)$ masses to the
$D^{*}(2010)\bar{D}^*(2010)$ and $D_1(2420)\bar{D}(1865)$ thresholds
respectively, we use the QCD sum rules (QCDSR) \cite{svz,rry,SNB}, to
study the two-point functions of the $D^*D^*$ molecule with $J^P=0^+$,
and the $D_1D$ molecule with $J^P=1^-$, to see if they can be interpreted as
the new observed resonances structures $Z_1^+(4050)$ and $Z_2^+(4250)$
respectively. Since they were observed in the $\pi^+\chi_{c1}$ channel,
the only quantum numbers that are known about them are $I^G=1^-$.

In  previous calculations, the QCDSR approach was used to study
the $X(3872)$ considered as a diquark-antidiquark state \cite{x3872}
and as a $D^*\bar{D}$ molecular state \cite{molecule},
 the $Z^+(4430)$ meson, considered as a $D^*\bar{D}_1$ molecular state
\cite{lee} and as tetraquark states \cite{bracco},
and the $Y$ mesons considered as molecular and
tetraquark states \cite{rapha} .
In some cases a very good agreement with the experimental mass was obtained.
The QCDSR approach was also used to study the existence of a $D_s\bar{D}^*$
molecule with $J^P=1^+$, that would decay into $J/\psi K^*\to J/\psi K\pi$
and, therefore, could be easily reconstructed \cite{molecule}.


Considering the $Z_1^+(4050)$ resonance structure as a
$D^{*+}\bar{D}^{*0}$ molecule with
$I^GJ^P=1^-0^+$, a possible current describing such state is given by:
\beq
j=(\bar{d}_a\gamma_\mu c_a)(\bar{c}_b\gamma^\mu u_b)
\;,
\label{field}
\enq
where $a$ and $b$ are color indices.

The sum rule is constructed from the two-point correlation function:
\beq
\Pi(q)=i\int d^4x ~e^{iq.x}\lag 0
|T[j(x)j^\dagger(0)]|0\rag.
\lb{2po}
\enq

On the OPE side, we work at leading order in $\alpha_s$ in the operators and
consider the contributions from condensates up to dimension eight.
The correlation function in the OPE side can be written as a
dispersion relation:
\beq
\Pi^{OPE}(q^2)=\int_{4m_c^2}^\infty ds {\rho^{OPE}(s)\over s-q^2}\;,
\lb{ope}
\enq
where $\rho^{OPE}(s)$ is given by the imaginary part of the
correlation function: $\pi \rho^{OPE}(s)=\mbox{Im}[\Pi^{OPE}(s)]$.

In the phenomenological side, we write a dispersion relation to the
correlation function in Eq.~(\ref{2po}):
\beq
\Pi^{phen}(q^2)=\int ds\, {\rho^{phen}(s)\over s-q^2}\,+\,\cdots\,,
\label{phen}
\enq
where $\rho^{phen}(s)$ is the spectral density and the dots represent
subtraction terms. The spectral density is described, as usual, as a single
sharp
pole representing the lowest resonance plus a smooth continuum representing
higher mass states:
\beqa
\rho^{phen}(s)&=&\lambda^2\delta(s-\mme^2) +\rho^{cont}(s)\,,
\label{den}
\enqa
where $\lambda$ gives the coupling of the current to the scalar meson $D^*
\bar{D}^*$:
\beq\label{eq: decay}
\lag 0 |
j|D^*D^*\rag =\lambda.
\label{lam}
\enq

For simplicity, it is
assumed that the continuum contribution to the spectral density,
$\rho^{cont}(s)$ in Eq.~(\ref{den}), vanishes bellow a certain continuum
threshold $s_0$. Above this threshold, it is assumed to be given by
the result obtained with the OPE. Therefore, one uses the ansatz \cite{io1}
\beq
\rho^{cont}(s)=\rho^{OPE}(s)\Theta(s-s_0)\;,
\enq

 After making a Borel transform to both sides of the sum rule, and
transferring the continuum contribution to the OPE side, the sum rules
for the scalar meson $Z_1^+$, considered as a scalar $D^*D^*$ molecule,
up to dimension-eight condensates, using factorization hypothesis, can
be written as:
\beq \lambda^2e^{-\mme^2/M^2}=\int_{4m_c^2}^{s_0}ds~
e^{-s/M^2}~\rho^{OPE}(s)\; +\Pi^{mix\qq}(M^2)\;, \lb{sr}
\label{sr1}
\enq
where
\beq
\rho^{OPE}(s)=\rho^{pert}(s)+\rh^{\qq}(s)
+\rh^{\lag G^2\rag}(s)+\rh^{mix}(s)+\rh^{\qq^2}(s)\;,
\lb{rhoeq}
\enq
with
\beqa\label{eq:pert}
&&\rho^{pert}(s)={3\over 2^{9} \pi^6}\int\limits_{\almi}^{\alma}
{d\al\over\alpha^3}
\int\limits_{\bemi}^{1-\al}{d\be\over\be^3}(1-\al-\be)
\left[(\al+\be)m_c^2-\al\be s\right]^4,
\nn\\
&&\rho^{\qq}(s)=-{3m_c\qq\over 2^{5}\pi^4}\int\limits_{\almi}^{\alma}
{d\al\over\al^2}
\int\limits_{\bemi}^{1-\al}{d\be\over\be}\left[(\al+\be)m_c^2-
\al\be s\right]^2,
\nn\\
&&\rho^{\lag G^2\rag}(s)={m_c^2\Gd\over2^{8}\pi^6}\int\limits_{\almi}^{
\alma}{d\al\over\alpha^3}\int\limits_{\bemi}^{1-\al}{d\be}(1-\al-\be)
\left[(\al+\be)m_c^2-\al\be s\right],
\nn\\
&&\rho^{mix}(s)=-{3m_cm_0^2\qq\over 2^{6}\pi^4}
\int\limits_{\almi}^{\alma}{d\al\over\al} [m_c^2-\al(1-\al)s],
\nn\\
&&\rho^{\qq^2}(s)={m_c^2\qq^2\over 4\pi^2}\sqrt{1-4m_c^2/s},
\label{dim6}
\enqa
where the integration limits are given by $\almi=({1-\sqrt{1-
4m_c^2/s})/2}$, $\alma=({1+\sqrt{1-4m_c^2/s})/2}$, $\bemi={\al
m_c^2/( s\al-m_c^2)}$, and we have used $\lag\bar{q}g\si.Gq\rag=m_0^2\qq$. 
We have neglected the contribution of the dimension-six
condensate $\langle g^3 G^3\rangle$, since it is assumed to be suppressed
by the loop factor $1/16\pi^2$. For completeness we have also included a part
of the dimension-8 condensate contributions
\beqa
&&\Pi^{mix\qq}(M^2)=-{m_c^2m_0^2\qq^2\over 8\pi^2}\int_0^1
d\al~e^{-m_c^2\over \al(1-\al)M^2}\left[1+{m_c^2
\over \al(1-\al)M^2}\right].
\label{dim8}
\enqa
One should note that a complete evaluation of the dimension-8 condensate 
contributions require more involved analysis including a nontrivial choice of 
the factorization assumption basis \cite{bnp}, which is beyond the scope of 
this calculation.

\begin{figure}[h]
\centerline{\epsfig{figure=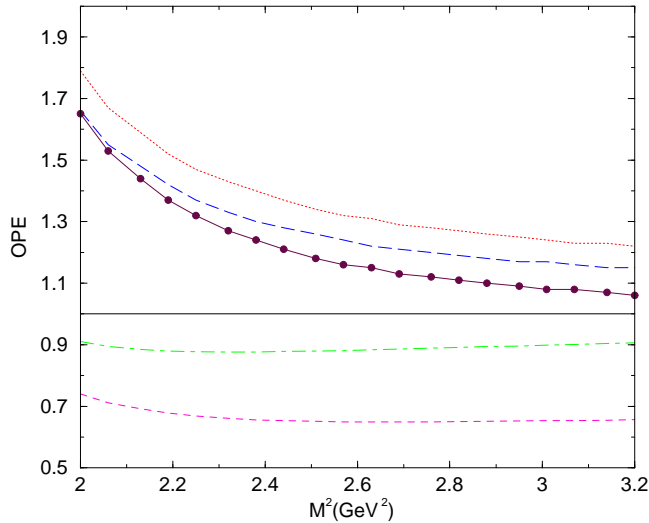,height=70mm}}
\caption{The OPE convergence for the $D^*D^*$ molecule in the region
$2.0 \leq M^2 \leq3.2~\GeV^2$ for $\sqrt{s_0} = 4.6$ GeV.  Comparing to the 
total contribution, we plot the relative contributions starting with
perturbative contribution (dashed line),
and each other line represents the relative contribution after adding of one
extra condensate in the expansion: + $\qq$ (dotted line), + $\langle g^2G^2
\rangle$ (long-dashed line), + $m_0^2\qq$ (dot-dashed line), + $\qq^2$
(line with circles), + $m_0^2\qq^2$ (solid line).}
\label{figconv}
\end{figure}

For a consistent comparison with the results obtained for the other molecular
states using the QCDSR approach, we  have considered here the same values 
used for the quark masses and condensates  as in 
refs.~\cite{lee,molecule,rapha,narpdg}:
$m_c(m_c)=(1.23\pm 0.05)\,\GeV $,
$\lag\bar{q}q\rag=\,-(0.23\pm0.03)^3\,\GeV^3$, $m_0^2=0.8\,\GeV^2$, 
$\lag g^2G^2\rag=0.88~\GeV^4$, where $g=\sqrt{4\pi\alpha_s}$.

To determine the Borel window, we analyse the OPE convergence and the pole 
contribution: the minimum value of the Borel mass is fixed by considering 
the convergence of the OPE, and the maximum value of the Borel mass is 
determined by imposing that the pole contribution must be bigger than the 
continuum contribution. To fix the continuum
threshold range we extract the mass from the sum rule, for a given $s_0$,
and accept such value of $s_0$ if the obtained mass is in the range 0.4 GeV
to 0.6 GeV smaller than $\sqrt{s_0}$. Using these criteria,
we  evaluate the sum rules in the Borel range $2.0 \leq M^2 \leq 3.5\GeV^2$,
and in the $s_0$ range $4.5\leq \sqrt{s_0} \leq4.7$ GeV.

From Fig.~\ref{figconv} we see that for $M^2\geq 2.5$ GeV$^2$
the contribution of the dimension-8 condensate is less than 20\% of the
sum of the other contributions. Using this fact as a criterion to establish
a reasonable OPE convergence, we  fix the lower
value of $M^2$ in the sum rule window as $M^2_{min} = 2.5$ GeV$^2$.

\begin{figure}[h]
\centerline{\epsfig{figure=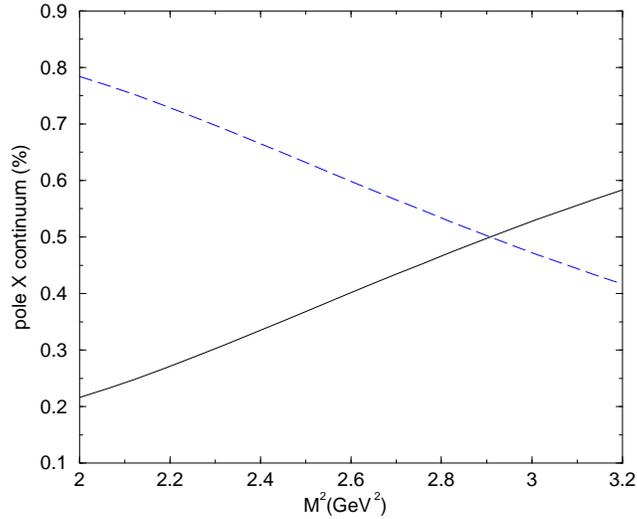,height=70mm}}
\caption{The dashed line shows the relative pole contribution (the
pole contribution divided by the total, pole plus continuum,
contribution) and the solid line shows the relative continuum
contribution for $\sqrt{s_0}=4.6~\GeV$.}
\label{figpvc}
\end{figure}

The comparison between pole and
continuum contributions for $\sqrt{s_0} = 4.6$ GeV is shown in
Fig.~\ref{figpvc}. From this figure we see that the pole contribution
is bigger than the continuum for $M^2\leq2.9~\GeV^2$.
The maximum value of $M^2$ for which this constraint is satisfied
depends on the value of $s_0$. The same analysis for the other values of the
continuum threshold gives $M^2 \leq 2.75$  GeV$^2$ for $\sqrt{s_0} =
4.5~\GeV$ and $M^2 \leq 3.1$  GeV$^2$ for $\sqrt{s_0} = 4.7~\GeV$.
In our numerical analysis, we shall then consider the range of $M^2$ values
from 2.5 $\GeV^2$ until the one allowed by the pole dominance criterion given
above.

\begin{figure}[h]
\centerline{\epsfig{figure=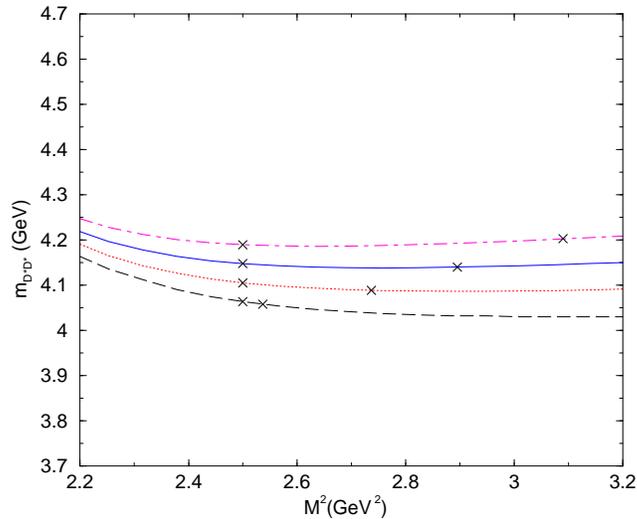,height=70mm}}
\caption{The $D^*D^*$ meson mass as a function of the sum rule parameter
($M^2$) for $\sqrt{s_0} =4.4$ GeV (long-dashed line), $\sqrt{s_0} =4.5$ GeV 
(dotted line), $\sqrt{s_0} =4.6$ GeV (solid
line) and $\sqrt{s_0} =4.7$ GeV (dot-dashed line). The crosses
indicate the upper and lower limits in the Borel region.}
\label{figmz1}
\end{figure}

To extract the mass $\mme$ we take the derivative of Eq.~(\ref{sr})
with respect to $1/M^2$, and divide the result by Eq.~(\ref{sr}).

In Fig.~\ref{figmz1}, we show the $D^*D^*$ meson mass, for different values
of $\sqrt{s_0}$, in the relevant sum rule window, with the upper and
lower validity limits indicated.  From this figure we see that
the results are very stable as a function of $M^2$.

Using the Borel window, for each value of $s_0$, to evaluate the mass of the
$D^*D^*$ meson and then varying the value of the continuum threshold in the
range $4.5\leq \sqrt{s_0} \leq4.7$
GeV, we get $\mme = (4.15\pm0.05)~\GeV$.

Because of the complex spectrum of the exotic states, some times lower 
continuum threshold values are favorable in order to completely eliminete the
continuum above the resonance state. Therefore, in Fig.~\ref{figmz1} we
also include the result for the $D^*D^*$ meson mass for $\sqrt{s_0}=4.4\GeV$.
We see that we get a very narrow Borel window, and for values of the 
continuum threshold smaller than 4.4 GeV there is no allowed Borel window. 
Considering then the continuum threshold in the range $4.4\leq \sqrt{s_0} 
\leq4.7$ GeV, we get $\mme = (4.13\pm0.07)~\GeV$.

To check the dependence of our results with the value of the
charm quark mass, we fix $\sqrt{s_0}=4.6~\GeV$ and vary the charm quark mass
in the range $m_c=(1.23\pm0.05)~\GeV$. Using $2.5\leq M^2\leq 2.9~\GeV^2$
we get $\mme = (4.15\pm0.07)~\GeV$.

Up to now we have taken the values of the quark-gluon mixed condensate
and the gluon condensate without allowing any uncertainties. While from
Fig.~\ref{figconv} we can see that a change in the gluon condensate value 
has little effect in our results, this is not the case for the
quark-gluon mixed condensate. Allowing $m_0^2$ to vary in the range
$m_0^2=(0.8\pm0.1)\GeV^2$ and fixing  $\sqrt{s_0}=4.6~\GeV$ we get
$\mme = (4.15\pm0.06)~\GeV$. Finally, assuming a possible violation of the
factorization hypothesis, one should multiply $\qq^2$ in Eqs.~(\ref{dim6}) and
(\ref{dim8}) by a factor $K$. Using  $K=2$, which means a violation of the
factorization hypothesis by a factor 2, and $\sqrt{s_0}=4.6~\GeV$ we get
$\mme = (4.12\pm0.02)~\GeV$. Therefore, taking into account the uncertainties
in the QCD parameters as discussed above we arrive at
\beq
\mme = \left(4.13\pm0.07^{+0.09~+0.08~+0.01}_{-0.05~-0.04~-0.03}\right)~\GeV,
\enq
where the first, second, third and forth errors come from the uncertainties
in $s_0,~m_c,~m_0^2$ and the factorization hypothesis respectively. Adding
the errors in quadrature we finally arrive at
\beq
\mme = (4.15\pm0.12)~\GeV,
\label{z1mass}
\enq
where the central value is around 130 MeV above the $D^*D^*(4020)$ threshold,
indicating the existence of  repulsive interactions between the two $D^*$ 
mesons.
Strong interactions effects might lead to repulsive interactions that
could result in a virtual state above the threshold. Therefore,
this structure may or may not indicate a resonance. However, considering the 
errors, it is compatible with the observed $Z_1^+(4050)$ resonance mass.

One can also deduce, from Eq.~(\ref{sr1}) the parameter 
$\lambda$ defined in Eq.~(\ref{lam}). We get:
\beq
\lambda = \left(4.20^{+0.68~+0.46~+0.45~+0.19}_{-0.88~-0.30~-0.19~+0.03}
\right)\times 10^{-2}~\GeV^5,
\label{la1}
\enq
where the first, second, third and forth errors come from the uncertainties
in $s_0,~m_c,~m_0^2$ and the factorization hypothesis respectively. Adding
the errors in quadrature we finally arrive at
$\lambda = (4.20\pm0.96)\times 10^{-2}~\GeV^5$.

Since to obtain the mass we have taken the derivative of the sum rule in
Eq.~(\ref{sr1}), it is important also to check if the convergence of the OPE 
and the pole contribution dominance are also satisfied for the derivative
sum rule.

\begin{figure}[h]
\centerline{\epsfig{figure=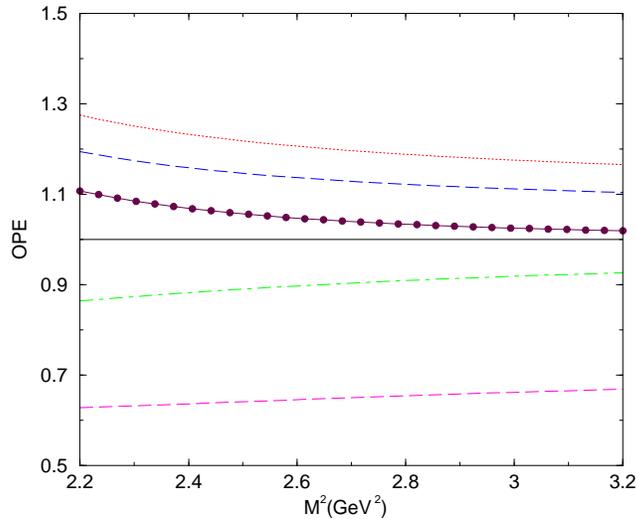,height=70mm}}
\caption{Same as Fig. 1 for the derivative of Eq.~(\ref{sr1}).}
\label{figconvderi}
\end{figure}
From Fig.~\ref{figconvderi} we see that the OPE convergence is even better
as from Fig.~\ref{figconv}. Therefore, it is correct to fix the lower
value of $M^2$ from the convergence of the original sum rule in  
Eq.~(\ref{sr1}). Regarding the pole contribution, we show in Table I
the values of $M^2$ for which the pole contribution is 50\% of the total
contribution, for each value of $\sqrt{s_0}$.

\begin{center}
\small{{\bf Table I:} Upper limits in the Borel window for the $D^*D^*$ 
molecule obtained from the derivative sum rule.}
\\
\begin{tabular}{|c|c|}  \hline
$\sqrt{s_0}~(\GeV)$ & $M^2_{max}(\GeV^2)$  \\
\hline
 4.5 & 2.42 \\
\hline
4.6 & 2.57 \\
\hline
4.7 & 2.72 \\
\hline
\end{tabular}\end{center}

From Table I we see that if we impose that both, the original sum rule and 
the derivative sum rule, should satisfy the OPE convergence and
the pole dominance  criteria, the Borel window exists only for $\sqrt{s_0}\geq
4.6\GeV$. Therefore, the result for the mass of the $D^*D^*$ molecule would be
even bigger than the result in Eq.~(\ref{z1mass}). This fact strongly
support our interpretation that the $D^{*+}\bar{D}^{*0}$ state is a 
virtual state and, probably, is not related with the recently observed 
$Z_1^+(4050)$. This result is in
disagreement with the findings of ref.~\cite{llz}.

Considering the $Z_2^+(4250)$ resonance structure as a $D_1D$ molecule
with $I^GJ^P=1^-1^-$, a possible current describing such state is given by:
\beq
j_\mu={i\over\sqrt{2}}\left[(\bar{d}_a\gamma_\mu\gamma_5 c_a)(\bar{c}_b
\gamma_5u_b)+(\bar{d}_a\gamma_5 c_a)(\bar{c}_b\gamma_\mu\gamma_5 u_b)
\right]\;.
\label{field2}
\enq
In this case, the two-point correlation function is given by:
\beq
\Pi_{\mu\nu}(q)=i\int d^4x ~e^{iq.x}\lag 0
|T[j_\mu(x)j_\nu^\dagger(0)]|0\rag.
\lb{2po2}
\enq
Since the current in Eq.~(\ref{field2}) is not conserved, we can write
the correlation function in Eq.~(\ref{2po2}) in terms of two independent
Lorentz structures:
\beq
\Pi_{\mu\nu}(q)=-\Pi_1(q^2)(g_{\mu\nu}-{q_\mu q_\nu
\over q^2})+\Pi_0(q^2){q_\mu q_\nu\over q^2}.
\lb{lorentz}
\enq
The two invariant functions, $\Pi_1$ and $\Pi_0$, appearing in
Eq.~(\ref{lorentz}), have respectively the quantum numbers of the spin 1
and 0 mesons. Therefore,
we choose to work with the Lorentz structure $g_{\mu\nu}$, since it
gets contributions only from the $1^{-}$ state.
The sum rule for the meson $Z_2^+$, considered as a vector $D_1D$ molecule,
in the Lorentz structure $g_{\mu\nu}$ can also be given by Eq.(\ref{sr1})
with:

\beqa\label{eq:pert2}
&&\rho^{pert}(s)={3\over 2^{12} \pi^6}\int\limits_{\almi}^{\alma}
{d\al\over\alpha^3}
\int\limits_{\bemi}^{1-\al}{d\be\over\be^3}(1-\al-\be)(1+\al+\be)
\left[(\al+\be)m_c^2-\al\be s\right]^4,
\nn\\
&&\rho^{\qq}(s)={3m_c\qq\over 2^{7}\pi^4}\int\limits_{\almi}^{\alma}
{d\al\over\al}
\int\limits_{\bemi}^{1-\al}{d\be\over\be^2}(1-\al-\be)\left[(\al+\be)m_c^2-
\al\be s\right]^2,
\nn\\
&&\rho^{\lag G^2\rag}(s)={\Gd\over2^{11}\pi^6}\int\limits_{\almi}^{\alma}
d\al\!\!\int\limits_{\bemi}^{1-\al}{d\be\over\be^2}\left[(\al+\be)m_c^2-\al
\be s\right]\left[{m_c^2(1-(\al+\be)^2)\over\be}-
{(1-2\al-2\be)\over\al}\left[(\al+\be)m_c^2-\al\be s\right]
\right],
\nn\\
&&\rho^{mix}(s)={3m_cm_0^2\qq\over 2^{8}\pi^4}
\int\limits_{\almi}^{\alma}{d\al} \int\limits_{\bemi}^{1-\al}{d\be\over\be^2}
(2\al+3\be)[(\al+\be)m_c^2-\al\be s],
\nn\\
&&\rho^{\qq^2}(s)=-{m_c^2\qq^2\over 2^4\pi^2}\sqrt{1-4m_c^2/s},
\nn\\
&&\Pi^{mix\qq}(M^2)=-{m_c^2m_0^2\qq^2\over 2^5\pi^2}\int_0^1
d\al~{e^{-m_c^2\over \al(1-\al)M^2}\over1-\al}\left[\al-{m_c^2
\over \al M^2}\right].
\label{dim82}
\enqa

\begin{figure}[h]
\centerline{\epsfig{figure=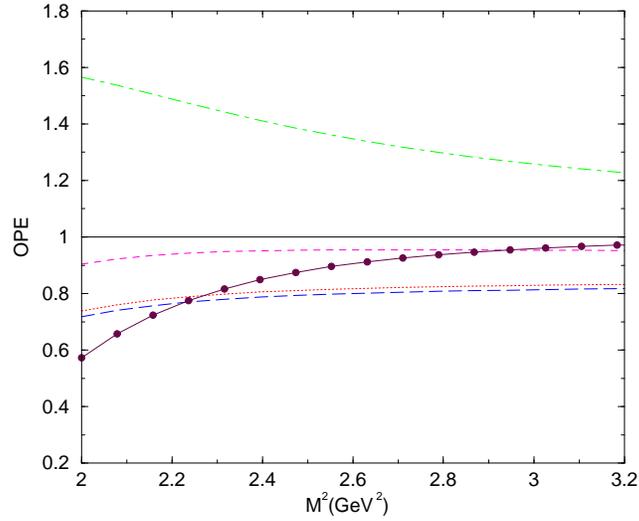,height=70mm}}
\caption{Same as Fig. 1 for the $D_1D$ molecule.}
\label{figopez2}
\end{figure}

In this case,
from Fig.~\ref{figopez2} we see that we obtain a reasonable OPE
convergence for $M^2\geq 2.4$ GeV$^2$. Therefore, we  fix the lower
value of $M^2$ in the sum rule window as $M^2_{min} = 2.4$ GeV$^2$.
The OPE convergence obtained from the derivative sum rule is better
than the OPE convergence from the original sum rule. From the derivative 
sum rule we get a good OPE convergence for $M^2\geq 2.2$ GeV$^2$.

The upper limits for $M^2$ for each value of $\sqrt{s_0}$ are given in
Table II, for the original sum rule and for the derivative sum rule.

\begin{center}
\small{{\bf Table II:} Upper limits in the Borel window for $D_1D$ molecule
with $J^P=1^-$.}
\\
\begin{tabular}{|c|c|c|}  \hline
$\sqrt{s_0}~(\GeV)$ & $M^2_{max}(\GeV^2)$ (sum rule in Eq.(\ref{sr1}))& 
$M^2_{max}(\GeV^2)$ (derivative sum rule)  \\
\hline
 4.5 & 2.56 & 2.25 \\
\hline
4.6 & 2.73 & 2.42 \\
\hline
4.7 & 2.90 & 2.58\\
\hline
\end{tabular}\end{center}

Again we see that the upper limits in the Borel window imposed by the
derivative sum rule are smaller than the ones obtined with the original sum 
rule, and this would restrict the range of values allowed for the
continuum threshold. However, we will allow a small violation in the 50\%
pole contribution criterion in the derivative sum rule, and we will work
in the Borel window allowed by the original sum rule. 

\begin{figure}[h]
\centerline{\epsfig{figure=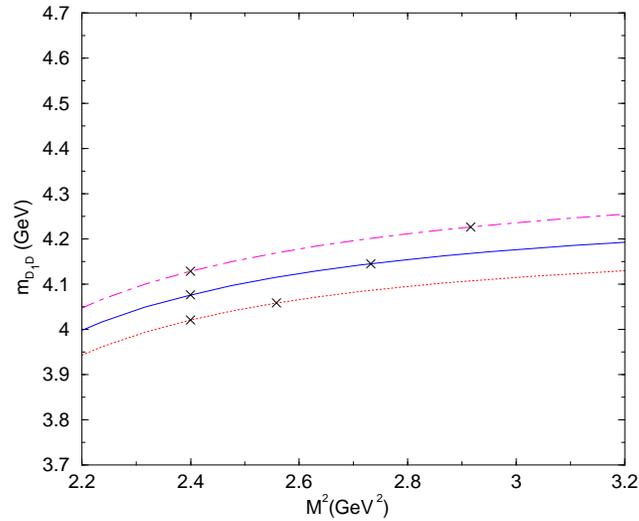,height=70mm}}
\caption{The $D_1D$ meson mass as a function of the sum rule parameter
($M^2$) for $\sqrt{s_0} =4.5$ GeV (dotted line), $\sqrt{s_0} =4.6$ GeV (solid
line) and $\sqrt{s_0} =4.7$ GeV (dot-dashed line). The crosses
indicate  the upper and lower limits in the Borel region.}
\label{figmz2}
\end{figure}

In the case of the $D_1D$ molecule we get a worse Borel stability than
for the $D^*D^*$, in the allowed sum rule window, as a function
of $M^2$, as can be seen by  Fig.~\ref{figmz2}. 

In Fig.~\ref{figmz2}, we show the $D_1D$ meson mass, for different values
of $\sqrt{s_0}$, in the relevant sum rule window, with the upper validity
limits indicated.  From this figure we see that
the results are not very stable as a function of $M^2$.

Using the value of the continuum threshold in the range $4.5\leq \sqrt{s_0} 
\leq4.7$ GeV, and varying $m_c,~m_0^2$ and $K$ as discussed above, we get
\beq
\mmo = (4.10^{+0.12~+0.05~+0.01~+0.28}_{-0.08~-0.05~-0.09~-0.02})~\GeV,
\enq
where the first, second, third and forth errors come from the uncertainties
in $s_0,~m_c,~m_0^2$ and the factorization hypothesis respectively. Adding
the errors in quadrature we finally arrive at
\beq
\mmo = (4.19\pm0.22)~\GeV,
\label{z2mass}
\enq
where the central value is around 100 MeV below the $D_1D(4285)$ threshold, 
and around 60 MeV smaller than the mass of the $Z_2^+(4250)$ resonance 
structure. Therefore, in this
case, there is an atractive interaction between the mesons $D_1$ and $D$
which can lead to the molecular state discussed above. Considering the
uncertainties in Eq.~(\ref{z2mass}), and the width
of the $Z_2^+(4250)$ resonance structure: $\Gamma_2=177^{+54+316}_{-39-~61}~
\MeV$, it seems to us that it is possible to describe this structure as a
$D_1D$ molecular state with $I^GJ^P=1^-1^-$ quantum numbers. However,
considering the uncertainties,  the result in Eq.~(\ref{z2mass}) is 
also compatible with the observed $Z_1^+(4050)$ resonance mass. Therefore,
no definite conclusion can be drawn for this state from the present
analysis.

For the value of the parameter $\lambda$ defined in Eq.~(\ref{lam}) we get:
\beq
\lambda_{Z_2} = \left(1.60^{+0.40~+0.30~+0.04~+0.66}_{-0.30~-0.30~-0.19~-0.10}
\right)\times 10^{-2}~\GeV^5,
\label{la2}
\enq
where the first, second, third and forth errors come from the uncertainties
in $s_0,~m_c,~m_0^2$ and the factorization hypothesis respectively. Adding
the errors in quadrature we finally arrive at
$\lambda_{Z_2} = (1.88\pm0.77)\times 10^{-2}~\GeV^5$.

In conclusion, we have presented a QCDSR analysis of the two-point
function for possible  $D^*\bar{D}^*$  and $D_1\bar{D}$ molecular states with
$I^GJ^P=1^-0^+$ and $I^GJ^P=1^-1^-$ respectively. 
For the $D^*\bar{D}^*$ molecule with $I^GJ^P=1^-0^+$ we got a mass around 
130 MeV above the $D^*D^*$ threshold, and around 100 MeV above the observed
$Z_1^+(4050)$ mass. In the case of the $D_1\bar{D}$ molecule with $I^GJ^P=
1^-1^-$  we got a mass around 100 MeV below the $D_1D$ threshold, and around 
60 MeV smaller than the observed $Z_2^+(4250)$ mass. In ref.~\cite{width}
it was found that the inclusion of the width, in the phenomenological side
of the sum rule, increases the obtained mass for molecular states. This means 
that the introduction of the width in our calculation, will increase the mass 
of the  $D^*\bar{D}^*$ and $D_1\bar{D}$ molecules.  As a result, the mass of 
the $D_1\bar{D}$ molecule will be closer to the
observed $Z^+(4250)$ mass, and the mass of the $D^*\bar{D}^*$ molecule will 
be far  from the $Z^+(4050)$ mass. Therefore, we conclude that
the $D^{*+}\bar{D}^{*0}$ state is probably a virtual state that is not
related with the $Z_1^+(4050)$ resonance-like structure recently observed
by the Belle Coll. \cite{belle3}. Considering the fact that the
$D^*D^*$ threshold (4020) is so close to the $Z_1^+(4050)$ mass and that
the $\eta^{\prime\prime}_c(3^1S_0)$ mass is predicted to be around 4050 MeV
\cite{god}, it is probable that the $Z_1^+(4050)$ is only a threshold
effect \cite{god}.  Another possibility is that the $D^*\bar{D}^*$  are in a 
relative p-wave state.  Such configuration will lead to a $J^P=1^-$ state as 
the naive s-wave decay of $\pi^+ \chi_{c1}$ would imply.  
In the case of the $D_1\bar{D}$ state, although from the present analysis
its mass is compatible with both, the $Z_1^+(4050)$ and the $Z_2^+(4250)$ 
resonance-like structures, in ref.~\cite{width} it was shown that
considering a width $\Gamma=60\MeV$ and $\sqrt{s_0}=(4.55\pm0.05)\GeV$
the mass obtained for the $D_1\bar{D}$ state is $m_{D_1{D}}=(4.27\pm0.03)
\GeV$, in an excelent agreement with the $Z_2^+(4250)$ 
mass. Therefore, their conclusion is that it is possible  to describe the 
$Z_2^+(4250)$ resonance structure as a $D_1D$ molecular state 
with $I^GJ^P=1^-1^-$ quantum numbers.

\section*{Acknowledgements}

This work has been partly supported by FAPESP and CNPq-Brazil and
by the Korea Research Foundation KRF-2006-C00011.


\end{document}